\documentclass[conference]{IEEEtran}
\IEEEoverridecommandlockouts
\usepackage{cite}
\usepackage{amsmath,amssymb,amsfonts}
\usepackage{algorithmic}
\usepackage{graphicx}
\usepackage{textcomp}
\usepackage{xcolor}
\def\BibTeX{{\rm B\kern-.05em{\sc i\kern-.025em b}\kern-.08em
    T\kern-.1667em\lower.7ex\hbox{E}\kern-.125emX}}
\begin{document}

\title{Plasma Light As Diagnostic For Wakefields Driven By Developing Self-Modulation Of A Long Particle Bunch\\
}

\author{\IEEEauthorblockN{Patric Muggli}
\IEEEauthorblockA{
\textit{Max Planck Institute for Physics}\\
Munich, Germany \\
muggli@mpp.mpg.de}
\and
\IEEEauthorblockN{Michele Bergamaschi}
\IEEEauthorblockA{
\textit{Max Planck Institute for Physics}\\
Munich, Germany \\
michele.bergamaschi@cern.ch}
\and
\IEEEauthorblockN{Jan Pucek}
\IEEEauthorblockA{
\textit{Max Planck Institute for Physics}\\
Munich, Germany \\
jan.pucek@cern.ch}
\and
\IEEEauthorblockN{Daniel Easton}
\IEEEauthorblockA{
\textit{Glenworth Associates Ltd} \\
St John's Innovation Centre\\
Cambridge, UK\\
daniel.easton@gw-associates.co.uk}
\and
\IEEEauthorblockN{Justin Pisani}
\IEEEauthorblockA{\textit{Glenworth Associates Ltd} \\
St John's Innovation Centre\\
Cambridge, UK\\
justin.pisani@gw-associates.co.uk}
\and
\IEEEauthorblockN{Jim Uncles}
\IEEEauthorblockA{\textit{Wright Design Ltd} \\
7 Wellington Court\\
Cambridge, UK\\
jim.uncles@wrightdesign.net}
}

\maketitle

\begin{abstract}
We outline plans to use plasma light emitted as atomic lines radiation as a diagnostic for wakefields driven in plasma by a self-modulating, long proton bunch. %
This diagnostic is built into the design of a new vapor/plasma source that will also allow for imposing a plasma density step of various height at various locations. %
\textcolor{black}{Such a step of a few percent in relative density placed at a location where the self-modulation process grows is predicted by numerical simulations to make the bunch drive wakefields with GV/m amplitude over hundreds of meters of plasma.}
\end{abstract}


\section{Introduction}
In particle accelerators based on plasma~\cite{bib:tajima,bib:chen}, the accelerating and focusing fields are sustained by a periodic perturbation in the density of plasma electrons (n$_{e0}$). %
Unlike in RF accelerators, where the accelerating structure is fabricated, is excited by an RF power source, and is fixed in the laboratory reference frame, in accelerators based on plasma, a driver, intense laser pulse or relativistic particle bunch, forms the structure as it travels along the plasma. %
The structure is attached to, and travels with the driver at nearly the speed of light. %
Energy is extracted from the driver by the plasma that sustains the fields, i.e., the wakefields. %
A fraction of the energy stored in wakefields and in oscillation of plasma electrons can be extracted by a witness bunch~\cite{bib:litos}. %
The structure and the wakefields cannot be characterized prior to the excitation event, and it is difficult to directly measure the amplitude of the accelerating field. %
Difficulties arise from the frequency of the wakefields that usually exceeds 100\,GHz ($\propto$~n$_{e0}^{1/2}$) and from their transverse extent ($\propto$~n$_{e0}^{-1/2}$) that is thus smaller than 1\,mm. % 
These parameters are chosen for the amplitude of the accelerating field ($\propto$~n$_{e0}^{1/2}$) to exceed 1\,GV/m~\cite{bib:dawson}. %
The amplitude of the accelerating field is therefore usually deduced from other measured quantities. %
For example, the energy gain per particle of a witness bunch can be divided by the plasma length and the charge of the witness particle to obtain an average (over the plasma length and energy distribution) accelerating field. %
It is usually lower than the amplitude of the field. %
Transverse imaging~\cite{bib:transverse} can yield the relative perturbation in the density of plasma electrons. %
The accelerating field value can then be calculated as the same relative fraction of the wavebreaking field amplitude the plasma can sustain~\cite{bib:dawson}. %
Longitudinal interferometry~\cite{LongInterferometry} can yield the same information, though averaged over the interaction length between the probe laser pulse and the wakefields. %

These measurements require acceleration events, electrons either from the plasma trapped by the wave or externally injected, or sophisticated laser pulse and optical systems. %

%%%%%%%%%%%%%%%%%
\section{Plasma Light Diagnostic}

We consider here a diagnostic for the relative amplitude of wakefields based on the fact that wakefields are energy deposited in the plasma by the driver, energy that must dissipate at least when the plasma decays. %
The system returns to a state similar to that before the event (i.e., neutral gas/vapor, at low repetition rate), except perhaps at a higher temperature. %
The energy of the wakefields, stored in electromagnetic fields and kinetic energy of oscillating plasma electrons, transforms into heat and radiation through collisions. % and into radiation, itself possibly also finally converted into heat. %
Timescales for the various phenomena vary greatly, from picoseconds for the period of oscillation of the wakefields, to microseconds for plasma decay~\cite{bib:spencer}. %bib:desynature}. %

The dissipation of energy is an intricate phenomenon and involves multifarious processes, as was described for example in its early stage of evolution~\cite{bib:plasmaSLAC}. %
Many of the processes involve collisions leading to exchange of momentum and energy, but also excitation and de-excitation of bound electrons from ions and neutrals of the system, and recombination. %
These atomic processes lead to emission of radiations at discrete (atomic) wavelengths and of continuous bremsstrahlung radiation in high-density plasmas. %
At the same time, the plasma with a (generally) small transverse extent (when compared to its length) expands radially during the energy dissipation process. %
Plasma particles interact with surrounding neutrals, for example creating additional ionized particles~\cite{bib:plasmaSLAC}.%

Determining the relative or absolute contributions of all these processes to energy dissipation is extremely challenging. %
However, it is reasonable to assume that over a small range of initial conditions (e.g., amplitude of wakefields, energy deposited, plasma densities), the relative contribution of these processes is essentially constant and proportional to the amount of energy deposited in the observation volume. %
The proportionality between the amount of light emitted in the 610\,nm line of neutral lithium (LiI) and the change in energy of an electron bunch driving wakefields in a plasma, as measured with a magnetic spectrometer, was demonstrated experimentally~\cite{bib:slacplasmalight}. %does the DESY manuscript show something else?
%We note here that in
In AWAKE~\cite{bib:muggliready} %, it is not practical to measure small energy loss ($\sim$10\,GeV) by the 400\,GeV proton bunch expected in experiments. %
the protons of the drive bunch have an energy of 400\,GeV and it is not practical to measure small energy loss or gain  on the order of only 10\,GeV expected over the 10\,m-long plasma. %  
%%%%%%%%%%%%%

\begin{figure}
\centering
\includegraphics[width=0.9\columnwidth]{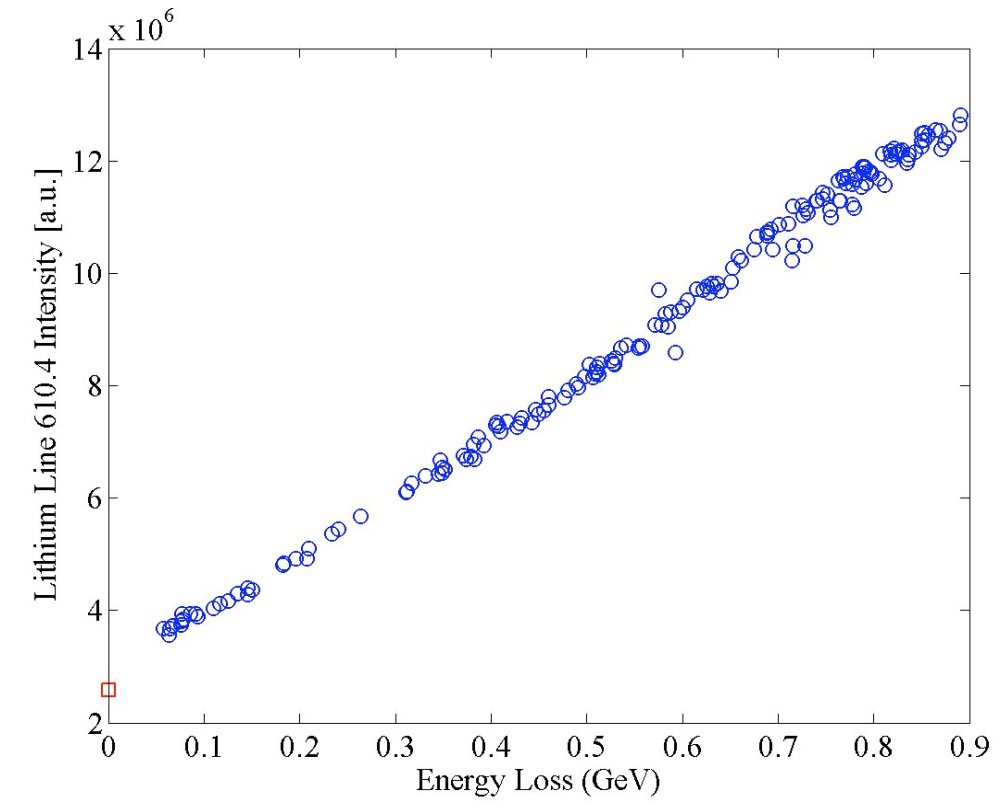}
\caption{\label{fig:frog}
Result of the measurement of the time-integrated light emitted by neutral lithium atoms measured with a gated CCD camera (time-integrated) as a function of the average energy loss by a short electron bunch in a plasma (blue symbols). %
The red square is the signal without plasma and corresponds to the optical transition radiation emitted by the bunch when entering the screen. %
Energy loss (average over the bunch particles) measured with a beam position monitor placed after the plasma in a dispersive region of the beamline. %
Variations obtained by changing the duration of the bunch. % 
Plasma electron density 10$^{17}$\,cm$^{-3}$. %
From~\cite{bib:slacplasmalight}. %
}
\end{figure}

%%%%%%%%%%%%%
%The signal is thus proportional to the amount of energy deposited, itself proportional to the square of the amplitude of the wakefields left behind the drive bunch. %
We note that in the  experiment of Ref.~\cite{bib:slacplasmalight}, variation of energy loss was obtained by varying the length of a single short bunch in the nonlinear regime of plasma wakefields~\cite{nonlinearPWFA,mugglireview}. %
The measured variation in energy of the bunch is thus the result of a convolution between change in amplitude of the wakefields and change in the number of particles experiencing decelerating and accelerating fields. %
The amount of energy loss of $\le$1\,GeV on average per 28.5\,GeV electron ($\cong$3\,J of the 93\,J of the bunch with 2$\times10^{10}$ particles and in $\cong$30\,cm of plasma) was small when compared to the energy of the incoming bunch. %
The amplitude of the wakefields was therefore most likely constant along the plasma \textcolor{black}{since the energy loss (average $<$1\,GeV by 28.5\,GeV electrons) is small when compared to their initial energy}. %

\textcolor{black}{We note here that plasma light was collected off a metallic screen otherwise used for measurements of the transverse size of the bunch using optical transition radiation (OTR). %
Prompt OTR was separated from plasma light using a gated-intensified camera with nanosecond gate time. %
The results of Fig.~\ref{fig:frog} (blue symbols) were obtained by increasing the gate time to ten microseconds to capture plasma light emitted over the time scale of dissipation of the energy of the wakefields. %
The contribution of the OTR light to the total signal is shown by the red symbol at the zero energy loss point obtained without plasma. %
}
%In previous experiments, the maximum average energy loss by $2\times10^{10}$ electrons was 1\,GeV. %
%The total energy loss was thus $\cong$3.2\,J. % 
%One can estimate the initial energy density of the wakefields, assuming this energy was deposited uniformly over the 10\,cm of the plasma length and over a radius of one cold plasma skin depth $c/\omega_{pe}\cong10.6\,\mu m$, as 85\,GJ/m$^{-3}$. %

This plasma light diagnostic relies on the collection of light emitted and its absolute value, it thus depends on the optical set up and its alignment and response. %
We therefore propose to use it to measure the amount of energy locally deposited by a long proton bunch self-modulating along a plasma~\cite{bib:kumar,bib:awake}. %
Self-modulation (SM) starts either from noise or irregularities in the parameters of the bunch driving wakefields~\cite{bib:fabian}, or from seed wakefields~\cite{bib:fabian,bib:livio}. %
The amplitude of modulation of the bunch charge density and the amplitude of wakefields grow in concert~\cite{bib:kumar}, until full self-modulation of the long bunch transforms it into a train of microbunches and the amplitude of wakefields saturates~\cite{bib:marlenesat}. %
Along a plasma with constant density, the amplitude of wakefields slowly decreases with further propagation. %
However, numerical simulation results show that when applying a few percent up-step in relative plasma density, a few meters along the plasma (i.e, during growth of SM), wakefields can maintain an amplitude close to their saturation value over a much longer distance~\cite{bib:lotovstep}. %
This is of course a necessary condition for an externally-injected electron bunch to experience large energy gain~\cite{bib:lotov200gev,bib:blumenfeld,bib:caldwellshort}. %

%%%%%%%%%%%%%%%%%%
\section{Current Vapor/Plasma Source}

In AWAKE, we use a rubidium (Rb) vapor source developed by the Max Planck Institute for Physics in collaboration with Wright Design Ltd and Glenworth Associates Ltd to produce a neutral density with a uniformity $\delta n_{Rb}/n_{Rb}<$0.2\% between 10$^{14}$ to 10$^{15}$\,cm$^{-3}$ over a 10\,m length~\cite{bib:erdemsource1,bib:erdemsource2,bib:fabianinterf}. %
This is achieved by imposing a temperature with a comparable uniformity along the source. %
This is obtained by flowing a liquid from both ends of the source towards its middle. %
The uniformity of the plasma, created by laser, field-ionization of the vapor ($>$99.9\% RbI to RbII~\cite{bib:karl}), is the same as those of the vapor and temperature. %
We note here that plasma density uniformity is essential to maintain synchronicity between the accelerated bunch and the wakefields along the acceleration processs ~\cite{bib:mugglirun2})
, not for the SM process. %
In fact, SM occurs even in plasmas with density varying linearly by up to $\pm$20\% over 10\,m~\cite{bib:falk,bib:pablo}. %

\section{Preliminary Results}

\begin{figure}
\centering
\includegraphics[width=0.45\textwidth]{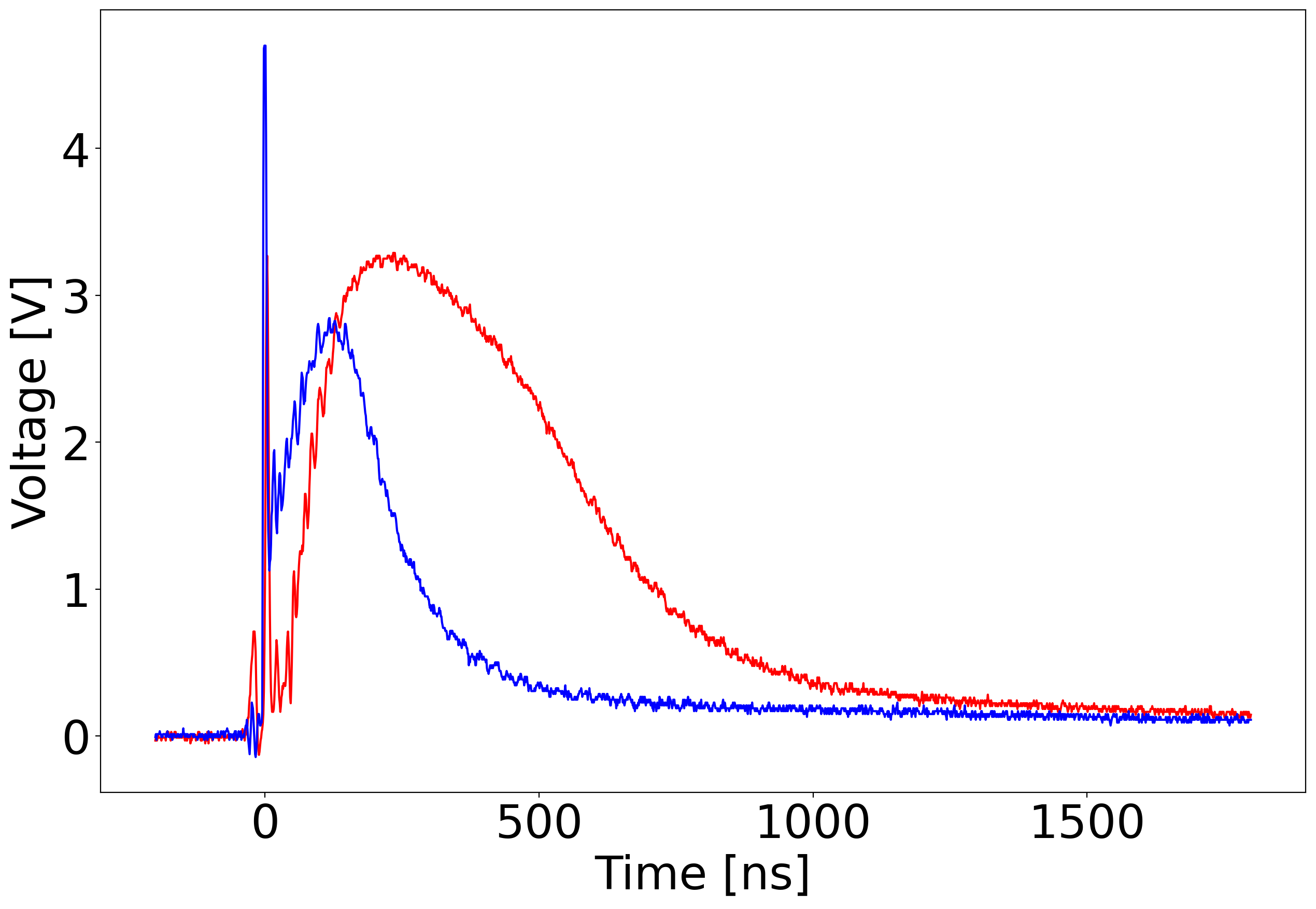}
\caption{\label{fig:traces}
Time history of the plasma light signal at the plasma entrance (red line) and exit (blue line) recorded by photo-multipliers. %
The very short peaks near t=0\,ns correspond to the ionizing laser pulse. %
Plasma density n$_{e0}$=1$\times$10$^{14}$\,cm$^{-3}$, bunch population 3$\times$10$^{11}$ protons, relativistic ionization seeding~\cite{bib:fabian}. %
}
\end{figure}

The source has two optical access ports that we routinely use to monitor and adjust the Rb vapor density~\cite{bib:fabianinterf}. %
These can also be used to record plasma light signals during events with plasma and/or proton bunch. %
Figure~\ref{fig:traces} shows typical plasma light signals, recorded at the very beginning and end of the vapor source. %
Plasma light occurs at wavelengths of Rb atomic transitions. %
The signal consists of these atomic lines filtered by the various elements of the system (spectral transmission by optical fibers, spectral response of the  photo-multiplier tubes, etc.), which peak in the visible optical range. %
The traces show a short initial signal (t$\cong$0\,ns) corresponding to light from the ionizing laser pulse scattered into collection optics by Rb neutral atoms. %
This signal is present without proton bunch and shows the response time of the photo-multiplier system ($\approx$\,1ns). %
On that time scale, it corresponds to the time at, and during which the proton bunch deposits energy (i.e., drives wakefields) into the plasma. %
Plasma light signals start at different times, peak with different values and at different times at the two locations, and return to approximately zero only after approximately 1\,\textmu s. %
This means that collisions between energetic plasma electrons and Rb neutral atoms (and ions), and thus also energy dissipation, still occur on that time scale, even though wakefields themselves may have a much shorter life time. %
Presumably, relaxation of the vapor to its initial state in the vapor source takes a longer time. %
%%%%%%%%%%%%%%%%%%%

%%%%%%%%%%%%%%%%%%%
Such measurements will allow to make measurements of relative plasma light signals with various relative heights and positions of the density step in order to optimize its effect on SM development and on the amplitude of wakefields (or associated potential or energy deposition). %
%%%%%%%%%%%%%%%%%%
\section{New Vapor/Plasma Source}

\begin{figure*}
\centering
\includegraphics[width=0.9\textwidth]{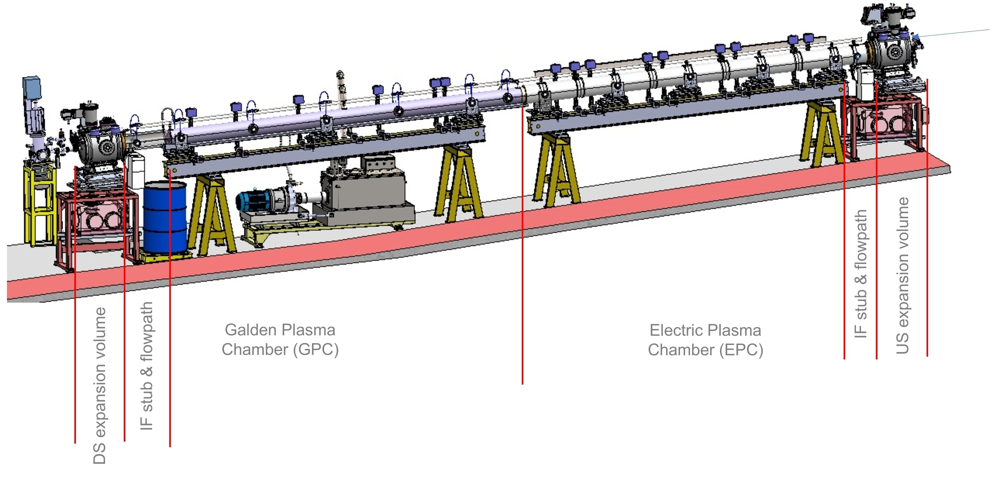}
\caption{\label{fig:VaporSourceCAD}
CAD drawing of the new vapor/plasma source including electrically heated section to allow for a density step to be imposed, three access ports to allow for measuring the two vapor densities, and ten access ports to measure plasma light signals. %
Beams propagate from right to left, the electrically-heated first 4.5\,m are labeled EPC (Electric Plasma Chamber). %
}
\end{figure*}

In the new source (see Fig.~\ref{fig:VaporSourceCAD}, also developed by the Max Planck Institute for Physics in collaboration with Wright Design Ltd and Glenworth Associates Ltd), we include three ports to measure vapor density~\cite{bib:fabianinterf}, and ten ports to measure plasma light signals along the plasma and SM development. %
%%%%%%%%%%%%%%%%%%%

%%%%%%%%%%%%%%%%%%%

Over the first 4.5\,m, the new source is heated by nine, electrically-powered, 50\,cm-long sections. %
The last 5.5\,m uses the same fluid heating scheme as the previous source. %
The temperature of the first eight heaters (the ninth one is set at the same temperature as that of the fluid) can be individually adjusted to generate a temperature and thus density step, with 0-10\% relative height. % 
%something about width(DT)?

\begin{figure}
\centering
\includegraphics[width=0.9\columnwidth]{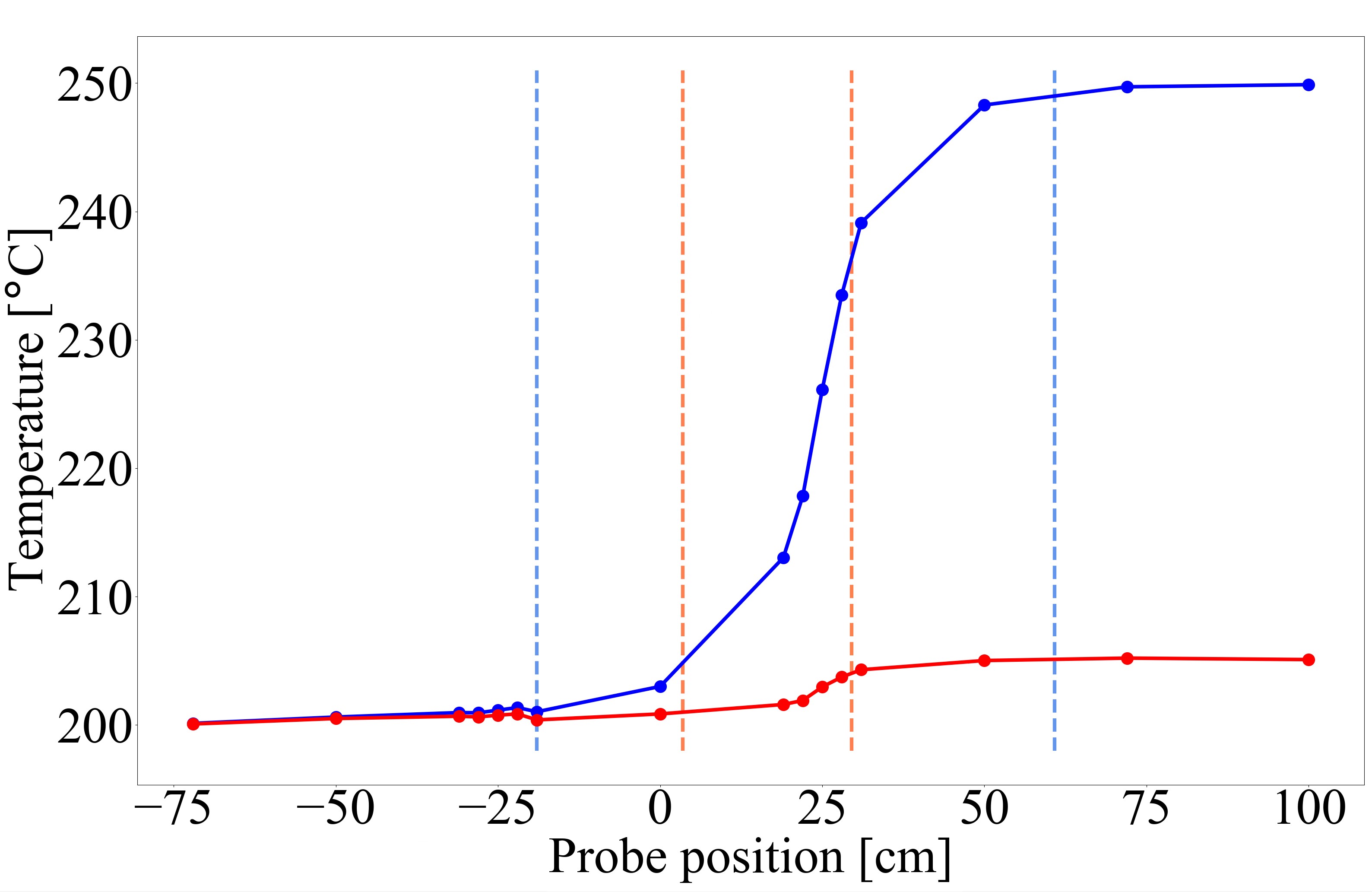}
\caption{\label{fig:HPP}
Temperature across two electrical heaters for two heights of the temperature step: $\cong$1\% (red symbols) and $\cong$10\% (blue symbols) measured on a prototype of the EPC with five heaters. %
Lines connect the measured points. %
The width of the step is about 25\,cm for the lower step (orange vertical lines) and about 80\,cm for the higher step (cyan vertical lines). %
}
\end{figure}

We built a prototype of the electrically heated section to test the ability to impose a temperature step. %
Figure~\ref{fig:HPP} shows that temperature steps of relative height 1 to 10\% can be imposed (above $\cong$200\textdegree C). %
The figure also shows that the width of the step is larger for the higher step: $\cong$80\,cm (201 to 249\textdegree C) rather than $\cong$25\,cm (201 to 204\textdegree C) for the lower step. %
However, numerical simulation results show that the effect of the step is weakly dependent on the step width~\cite{bib:lotovstep}. % 

\begin{figure}
\centering
\includegraphics[width=0.9\columnwidth]{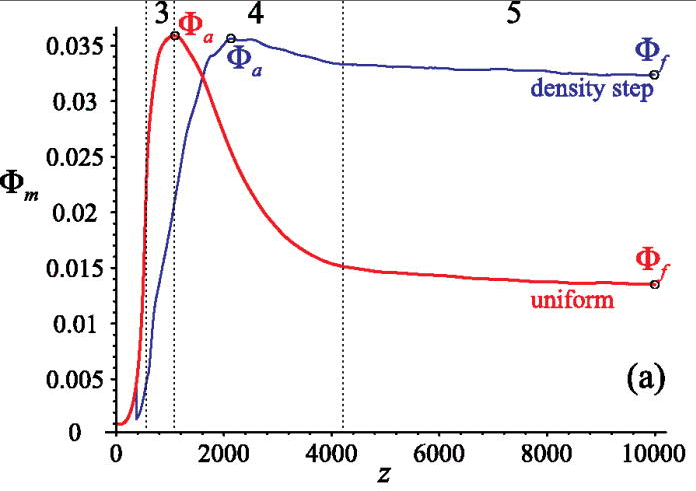}
\caption{\label{fig:SMpotential}
Result of numerical simulations showing the effect of an optimized (position and relative height) density step on the maximum potential $\Phi_m$ of the wakefields driven along the plasma by a positron bunch (for simulation speedup). %
Potentials are normalized to $c/\omega_{pe}E_{WB}$, and the distance to $c/\omega_{pe}$ the skin depth of the cold plasma, where $E_{WB}$ is the wavebreaking field in the cold plasma~\cite{bib:dawson}. %
The $\Phi_a$ and $\Phi_f$ are the values of maximum  potential and potential at the end of the plasma. %
The numbers refers to different regions of the SM growth (see~\cite{bib:lotovstep}). %
The density step yielding the largest $\Phi_f$ is located at $z=400$ (blue line). %
In this case the value of $\Phi_f$ is about two times larger with than without the density step. %
%The green dots suggest the possible results of plasma light measurements at ten locations. % 
From~\cite{bib:lotovstep}. %
}
\end{figure}

Figure~\ref{fig:SMpotential} shows an example of numerical simulation results~\cite{bib:lotovstep} showing the effect of a plasma density step with optimum parameters on the maximum of the potential of the wakefields with and without the density step. %
%%%%%%%%%%%%%%%%%%%%%%%%%%
%%%%%%%%%%%%%%%%%%%%%%%%%%
For the plasma light diagnostic, the challenge is to calibrate the ten measurements to obtain the information suggested on Fig.~\ref{fig:SMpotential}, i.e., first, growth  and saturation of SM ($\le$5\,m), and second drop or the maintaining of the amplitude of wakefields ($>$5\,m) with and without the optimized density step. %
%%%%%%%%%%%%%%%%%%%%%%%%%%
%%%%%%%%%%%%%%%%%%%%%%%%%%
Meaningful comparison of the signals at various locations along the plasma depends on the light collection efficiency (alignment) and calibration of the ten light measurements systems (optical couplers and fibers, photomultipliers). % 

Initial measurements suggest that alignment (maximization of the amount of light collected) and calibration could be performed using light from the laser pulse scattering on the atoms of a low density Rb vapor. %
The central wavelength of the laser pulse (780\,nm) is very close those of the D1 (781\,nm) and D2 (794\,nm) lines of Rb atoms, which, measurements show, accounts for about half the amount of plasma light collected. %
Therefore, a good relative calibration could be obtained from these measurements. %
Calibrations of the broadband response could be obtained from a black body source. %  
At low vapor density and low energy of the laser pulse, a low ionization fraction can be expected, thus avoiding the decrease of laser signal associated with energy or photon depletion of the laser pulse expected along propagation when significant ionization occurs~\cite{bib:gabor}. % 

Besides information about the effect of the density step, additional information could be obtained regarding optimization of the amplitude of wakefields using seed electron bunch and proton bunch parameters (e.g., charge), as suggested in one of our recent publications~\cite{bib:livio}. %
Plasma light signals could also show that seeding not only makes the phase or timing of SM reproducible~\cite{bib:fabian,bib:livio} along the bunch, but in addition that the amplitude of wakefields at all ten locations also becomes reproducible. %

In addition, we measured the effect of linear density gradients on the proton bunch itself~\cite{bib:falk,bib:pablo}, but not on the amplitude of the wakefields. %
Plasma light measurements could confirm that higher energy gains by side-injected electrons obtained with small positive linear density gradient (5\% over 10\,m)~\cite{bib:royal} were indeed the result of larger amplitude of wakefields. % and not of different capture location along the plasma. %

The effect of the density step will also be investigated by measuring energy gain by externally injected test electrons (as in~\cite{bib:AWAKEaccel} and ~\cite{bib:royal}), as well as by measuring the size and shape of the bunch halo on time-integrated images of the bunch (as in~\cite{bib:marlene}). %
%%%%%%%%%%%%%%%%%%
\section{Summary}

In AWAKE the driving of wakefields cannot be optimized by observing energy loss by the drive proton bunch, at least not over short plasma distance. %
%Indeed, measuring an energy loss per 400\,GeV proton on the order of 10\,GeV ($\approx$1\,GV/m over 10\,m) is extremely challenging. %
Other measurement for example for %energy loss and 
the amplitude of wakefields must be developed. %
Plasma light measurements may provide such an opportunity, as was demonstrated in~\cite{bib:slacplasmalight}. %
In addition, since the amplitude of the wakefields evolves significantly during SM, plasma light measurements may allow to measure, characterize and optimize this process in a self-modulator plasma, in order to maximize energy gain in a following accelerator plasma~\cite{bib:mugglirun2}, using experimental parameters~\cite{bib:livio} and a step in plasma density~\cite{bib:lotovstep}. %
Many characteristics of the SM process, previously inferred from characterization of particles exiting the plasma could be corroborated by measurements along the plasma itself, such as: growth~\cite{bib:karl} and saturation~\cite{bib:marlenesat} along the plasma, and reproducibility of the SM process~\cite{bib:fabian,bib:livio} not only in phase or timing, but also in amplitude. %
This last characteristic is of course essential for reproducible acceleration from event-to-event. %
Moreover, the effect of plasma density gradients on wakefields~\cite{bib:falk} and on energy gain by test electrons~\cite{bib:royal} could be attributed to changes in amplitude of wakefields. %
Plasma light could thus become a standard and simple diagnostic in plasma-based accelerators and beam-plasma interaction experiments to evidence any phenomenon that changes the kinetic energy of plasma electrons, including hosing and current filamentation instability~\cite{bib:cfi} interfering for example with the driving of wakefields~\cite{bib:nagaitsev}. %
\textcolor{black}{We note here that plasma light measurements have been used to measure energy transfer efficiency from the drive to the witness bunch of a plasma wakefield accelerator~\cite{boulton2022longitudinally}, supporting the value and validity of the diagnostic.}%

%%%%%%%%%%%%%%%%%%%%

\bibliography{sample.bib}{}
\bibliographystyle{IEEEtran}

\end{document}